\documentclass[preprint,aps]{revtex4}
\usepackage{epsfig} 
\usepackage{color} 
\input boxedeps.tex
\SetRokickiEPSFSpecial  
\HideDisplacementBoxes
\usepackage{amsmath}
\newcommand{\be}{\begin{equation}}
\newcommand{\eq}{\end{equation}}
\newcommand{\da}{\dagger}  
\newcommand{\perpto}[1]{\mathrm{Perp}(#1)}
\newcommand{\g}{\gamma}
\newcommand{\Tr}{{\rm \, Tr \!}}

\begin{document}
\mbox{SWAT/414}\hfil\\
\title{Small-$x$ Behaviour of Lightcone
Wavefunctions \\
in Transverse Lattice Gauge Theory} 

\author{J. Bratt${}^*$, S. Dalley${}^\da$, B. van de Sande${}^*$, 
and E. M. Watson${}^*$}
\affiliation{${}^*$Geneva College\\
         3200 College Ave., Beaver Falls, PA~~15010}
\affiliation{${}^\da$Department of Physics, University of Wales Swansea\\
Singleton Park, Swansea SA2 8PP, U.K.}

\begin{abstract}
We study  the behaviour of lightcone wavefunctions, in coarse transverse
lattice gauge theory, when one or more parton lightcone momenta
are small. This probes the limit of hadron structure at 
large and small Bjorken $x$.
Finite-energy boundary conditions 
on boundstates allow one to derive the analytic form of wavefunctions
in this region.
This  leads to simple, universal
predictions for the behaviour of quark generalised parton 
distributions near their endpoints. 
For the first few meson wavefunctions at large $N_c$,
we give the simplest  ansatz that
incorporates all the boundary conditions. 
\end{abstract}

\maketitle
\baselineskip .2in

\section{Introduction}
\label{intro}
Transverse lattice gauge theory \cite{bardeen,review}
is a promising approach for ab-initio
calculations of the hadronic light-cone wavefunctions. All 
successful calculations so far have been performed on coarse transverse
lattices with effective `colour-dielectric'
degrees of freedom, keeping the longitudinal
coordinates $x^- = (x^0 - x^3)/\sqrt{2}$ continuous. 
Although this has required some
parameters in the effective Hamiltonian to be fit phenomenologically,
it has the great computational advantages that 
hadrons are dominated by states containing few partons and 
there is a clean separation between the longitudinal and transverse 
dynamics.
In fact, in this case, it is not much
harder to analyse the dependence of wavefunctions on the longitudinal
coordinates than in $1+1$-dimensional gauge theories. While the full
wavefunction of a boundstate typically requires---even in 1+1 
dimensions---a numerical solution,
the behaviour when parton lightcone momenta
$k^+$ vanish can be largely understood analytically. 
The prototype example is $1+1$-dimensional large-$N_c$ QCD \cite{hoof},
where the behaviour of the meson wavefunction for vanishing quark
$k^+$ was found exactly. 
A similar analysis can be performed to some extent even in $3+1$-dimensional 
large-$N_c$ QCD \cite{ladder}. The key strategy in all cases is to demand
cancellation of singularities in the lightcone boundstate Schrodinger
equation when parton lightcone momenta vanish. Typically, wavefunctions
do not simply vanish in this high free-energy limit, as they would 
at large transverse momentum. Rather, there occur
`ladder' relations between states with different numbers of partons that
couple the analytic structure of the infinite tower of lightcone
wavefunctions.

In this paper, we analyse in a similar vein the corresponding 
lightcone boundstate Schrodinger
equation for coarse transverse lattice gauge theory at large-$N_c$. 
By demanding finiteness of the boundstate energy, we determine the
form of meson wavefunctions at small lightcone momentum.
These depend on two exponents,
$\alpha$ and $\beta$, whose values are independent
of all  details
of the wavefunctions, which encode the leading non-analytic behaviour.
In general, naive intuition based on the free lightcone kinetic energy 
is wrong. In particular, wavefunctions become more singular 
as more momenta vanish. Quark distributions diverge as $x \to 0$
and all Fock sectors contribute as  $x \to 1$.

In the next section, we 
briefly review  the colour-dielectric formulation of
transverse lattice gauge theory and set up the boundstate
Schrodinger equation to lowest non-trivial order of the colour-dielectric
expansion that is used in the following two sections. 
Sections~\ref{onelink} and \ref{analytic}
determine systematically the behaviour of
lightcone wavefunctions of mesons at small parton lightcone momentum.
At first, in section~\ref{onelink}, we use
the approximation where up to one gauge link-variable is
allowed in a state. Then, in section~\ref{analytic}, we relax 
this approximation
and generalise many of the results, giving a simple ansatz
that
incorporates all the boundary conditions.
Applications of the results to the behaviour of quark generalised parton
distributions
are given in Section~\ref{appl} and conclusions are discussed
in Section~\ref{last}.

\section{Review of Transverse Lattices}
\label{rev}
We introduce continuum light-cone co-ordinates $x^{\pm}= (x^0 \pm
 x^3)/\sqrt{2}$ and discretize the transverse coordinates ${\bf  x}=(x^1,x^2)$
on a square lattice of spacing $a$. Lorentz indices $\mu, \nu $ are
split into light-cone indices $\alpha,\beta \in \{+,-\}$
and transverse indices $r,s\in \{1,2\}$. 
Quark fields $\Psi(x^+,x^-,{\bf x})$ in the fundamental
representation and  longitudinal gauge potentials
 $A_{\alpha}(x^+,x^-,{\bf x})$ in the adjoint representation of
the gauge group $SU(N_c)$ are associated to the sites of the
 transverse
lattice. Flux link fields $M_r(x^+,x^-,{\bf x})$, which we choose to be in
a complex
  matrix representation of $SU(N_c)$,  are associated
with the directed link from ${\bf x}$ to ${\bf x} + a \hat{\bf
 r}$, where $\hat{\bf  r}$ is a unit vector in direction $r$. 
In the transverse continuum limit $a \to 0$, these would be related
to  transverse gauge potentials $A_{r}(x^+,x^-,{\bf x})$.
Subsequent analysis is done
to leading order of  the $1/N_c$ expansion, which allows one
to drop the site argument ${\bf x}$  of the fields \cite{morris,colour}.

\subsection{Lightcone Hamiltonian}
For finite transverse lattice spacing $a$, the Lagrangian may contain any
operator invariant under transverse lattice gauge symmetries, 
Poincar\'e symmetries
manifestly
preserved by the lattice cut-off, and renormalisable by dimensional
counting with respect to the continuum co-ordinates $x^{\alpha}$. 
One may control the subsequent proliferation of operators 
either by approaching the transverse continuum
limit, in which case only renormalisable operators with respect to 
co-ordinates ${\bf x}$ survive, or by `strong-coupling' methods on a 
coarse lattice, such as the colour-dielectric expansion \cite{colour}. 
All calculations to date
have used the colour-dielectric form \cite{review}
and it is this approach that we
analyse here. 
This involves fixing to the light-cone gauge $A_{-} =
0$, classically 
eliminating non-dynamical fields $A_+$, $\gamma^- \Psi$, 
then expanding the resulting
lightcone Hamiltonian in powers of the remaining dynamical fields
$M_r$, $\gamma^+ \Psi$. 
Truncation of
such an expansion is a valid approximation provided
wavefunctions of interest (typically those of the lightest
eigenstates) are dominated by few-body Fock states. 
This is achieved
by the choice of 
sufficiently large masses for the dynamical
fields.
When the remaining couplings in the truncated hamiltonian
are tuned to optimise symmetries broken by the lattice cutoff,
the choice of large masses corresponds to choice of 
a coarse lattice spacing $a$.

The Lagrangian density $L$ we consider, which was previously investigated in 
Ref.~\cite{mesons},
contains all allowed terms up to order  $M^4$ and 
$\overline{\Psi} M \Psi$.
(There are additional terms involving
multiple traces of products of $M$ fields that contribute only to
the glueball sector at large-$N_c$ \cite{glueballs}.)
\begin{eqnarray}
L & = &   \sum_{\alpha, \beta = +,-}
\sum_{r=1,2} 
-\frac{1}{2 G^2} \Tr \left\{ F^{\alpha \beta}F_{\alpha \beta} \right\}
 + \Tr\left\{[\left(\partial_{\alpha} + {\rm i} A_{\alpha} \right)
        M_r-  {\rm i} M_r   A_{\alpha}][{\rm h.c.}]\right\}
\nonumber \\
&& - \mu_{b}^2  \Tr\left\{M_r M_r^{\da}\right\} - V[M] 
 + {\rm i} \overline{\Psi} 
\g^{\alpha} (\partial_{\alpha} + {\rm i} A_{\alpha}) \Psi - \mu_f
\overline{\Psi}\Psi  
\nonumber\\
&& +  {\rm i} \kappa_a \left( \overline{\Psi} \g^{r} M_r \Psi 
- \overline{\Psi} \g^{r} M_{r}^{\da} 
\Psi \right)
+ \kappa_s \left( \overline{\Psi} M_r
 \Psi+\overline{\Psi} M_{r}^{\da}
 \Psi\right) 
\ , 
\label{ferlag}
\end{eqnarray}
where $F^{\alpha \beta}({\bf x})$ is the continuum field strength in the
$(x^0,x^3)$ plane and the link-field potential is 
\begin{eqnarray}
 V[M] & = & 
- \frac{\beta}{a^2 N_c} \sum_{r \neq s} 
  \Tr\left\{ M_{r} M_{s} M_{r}^{\da} M_{s}^{\da}
\right\} 
+ \frac{\lambda_1}{a^{2} N_c} \sum_r  
\Tr\left\{ M_r M_r^{\da}
M_r M_r^{\da} \right\} \nonumber \\
&& +  \frac{\lambda_2}{a^{2} N_c}\sum_r  
\Tr\left\{ M_r  M_r M_r^{\da} M_r^{\da} \right\} 
+  \frac{\lambda_4}{a^{2} N_c}  
\sum_{\sigma=\pm 2, \sigma^\prime = \pm 1}
        \Tr\left\{ 
M_\sigma^{\da} M_\sigma M_{\sigma^\prime}^{\da} M_{\sigma^\prime} \right\} 
 \; . \label{pot1}
\end{eqnarray}
We have defined $M_{r} = M_{-r}^{\dagger}$ and  hold
$ {G} \sqrt{N_c}$, $\kappa_a \sqrt{N_c}$, and $\kappa_s \sqrt{N_c}$
finite as $N_c \to \infty$.

In the chiral representation, 
\begin{equation}
         \Psi=\frac{1}{ 2^{1/4}}
\left(\begin{array}{c}u_+\\ v_+ \\ v_- \\ u_- \end{array}\right) 
\end{equation}
decomposes
into two-dimensional  
complex  spinors 
\begin{equation}
         u=\left(\begin{array}{c}u_+\\ u_-\end{array}\right) \quad
         v=\left(\begin{array}{c}v_+\\ v_-\end{array}\right) \ .
\end{equation}
The helicity
subscript $h = \pm$ denotes the sign of the eigenvalue of
$\gamma^5$.
Using Pauli matrices, we may then write various bilinears
\begin{eqnarray}
  \bar{\Psi}\Psi &=& -\frac{1}{\sqrt{2}}\left(u^\da \sigma^1 v+
                                  v^\da \sigma^1 u\right)\\
  \bar{\Psi}\gamma^+ \Psi &=& u^\da u\\
  \bar{\Psi}\gamma^- \Psi &=& v^\da v\\
  \bar{\Psi}\gamma^1 \Psi &=& \frac{1}{\sqrt{2}}\left(u^\da \sigma^3 v+
                                  v^\da \sigma^3 u\right)\\
  \bar{\Psi}\gamma^2 \Psi &=& \frac{\rm i}{\sqrt{2}}\left(v^\da u-
                                  u^\da v\right) \\
  \bar{\Psi}\gamma_5 \Psi &=& \frac{\rm i}{\sqrt{2}}\left(v^\da \sigma^2 u+
                                  u^\da \sigma^2 v\right) \; .
\end{eqnarray}
We also
 introduce the matrices $\zeta^\lambda$ which act on the 2-spinors:
\begin{equation}
\begin{array}{c|cccc}
  \lambda & 1 & 2 & -1 & -2\\
\hline
  \zeta^\lambda & \sigma^2 & -\sigma^1 & -\sigma^2 & \sigma^1 
\end{array} .
\end{equation}

In light-cone gauge $A_- = 0$,
$A_{+}$ and $v$ are non-dynamical (independent of light-cone
time $x^+$) and their 
constraint equations of motion 
are used to eliminate them at the classical level. The lightcone
Hamiltonian may then be obtained from the action (\ref{ferlag}) 
in the standard way in terms of
the remaining dynamical fields $u$, $M_{1}$, and $M_2$.
If we define
\begin{eqnarray}
    F & = & -  u + 
\sum_\lambda  M_\lambda \left( \frac{\kappa_s}{\mu_f} 
                    + {\rm i} \frac{\kappa_a}{\mu_f} \zeta^\lambda \right) u 
\label{xi} \  , \\
J^{+} &=& uu^{\dagger} + {\rm  i} \sum_{\lambda} 
M_{\lambda} \stackrel{\leftrightarrow}{\partial}_{-} 
M_{\lambda}^{\da} \label{current} \ ,
\end{eqnarray}
the lightcone Hamiltonian one finds is
\begin{eqnarray}
 P^-  &  = &  \int dx^-  \ 
   \frac{G^2}{ 4} \Tr\left\{ 
              J^{+} \frac{1}{({\rm i} \partial_{-})^{2}} J^{+} \right\}
            -\frac{G^2}{ 4 N_{c}} 
        \Tr\left\{ J^+  \right\} \frac{1}{ ({\rm i}\partial_{-})^{2} }
     \Tr\left\{ J^+ \right\}   \nonumber \\
&&  \ \ + \frac{\mu_{f}^{2}}{ 2}  
F^{\da} \frac{1}{ {\rm i} \partial_-}F  +    V[M]
\ . \label{ham} 
\end{eqnarray}
Under certain reasonable assumptions \cite{sdmesons}, 
the Hamiltonian (\ref{ham}) is a truncation of the most general  
Hamiltonian to order $M^4$ and $u^2 M$. It also contains some, but not
all, allowed operators at order $u^2 M^2$ and $ u^4 $; most 
importantly, it contains
those responsible for confinement \cite{bardeen}.

\subsection{Fock space}
Introducing the mode expansions
\begin{eqnarray}
 M_r(x^+=0,x^-)   &=&  
        \frac{1}{\sqrt{4 \pi }} \int_{0}^{\infty} \frac{dk^+}{ \sqrt{ k^+}}
        \left( a_{-r}(k^+)\, e^{ -{\rm i} k^+ x^-}  +  
        a^{\da}_r(k^+)\, e^{ {\rm i} k^+ x^-} \right)   \; , \nonumber \\
 u(x^+=0,x^-)   &=&  
        \frac{1}{\sqrt{2 \pi }} \int_{0}^{\infty} dk^+ 
        \left( b(k^+)\, e^{ -{\rm i} k^+ x^-}  +  
        \sigma^1 d^{\da}(k^+)\, e^{ {\rm i} k^+ x^-} \right)   \; ,
\end{eqnarray}
we have
\begin{eqnarray}
 \left[a_{\lambda,ij}(k^+), a_{\rho,kl}^{*}(\tilde{k}^+)\right] 
        & = & \delta_{ik}\, \delta_{jl}\, \delta_{\lambda \rho}\, 
        \,\delta(k^+-\tilde{k}^+) \; , \\
   \left[a_{\lambda,ij}(k^+),a_{\rho,kl}(\tilde{k}^+)\right] & = & 0 \; , \\
\left\{b_{h,i}(k^+), b_{h',j}^{*}(\tilde{k}^+)\right\} 
        & = & \delta_{ij}\,  \delta_{hh'}\, 
        \,\delta(k^+-\tilde{k}^+) \;, \\
   \left\{b_{h,i}(k^+), b_{h',j}(\tilde{k}^+)\right\} & = & 0 \; , 
\end{eqnarray}
where $i,j$ are colour indices,
$\lambda$ and $\rho \in \{ \pm 1, \pm 2\}$ denote the
orientations of link variables in the $(x^1,x^2)$ plane, 
and $\left(a_{\lambda,ij}\right)^{*} = 
(a_{\lambda}^{\dagger})_{ji}$.
The sequence of orientations $\{ \lambda_2, \cdots \lambda_{n-1}\}$ of link
variables together with the $P^+$ momentum fractions $x_a=k_a^+ / P^+$
and quark
helicities $h,h'$ are sufficient to encode the structure of Fock
states contributing to the boundstate at zero total transverse  momentum
${\bf P}= {\bf 0}$. 
Only gauge singlet combinations under residual gauge transformations in
$A_{-}=0$ gauge can contribute to states of finite energy \cite{bardeen}, so
a boundstate containing quarks at large $N_c$ 
decomposes in terms of  Fock states as
\begin{eqnarray}
|\psi(P^+)\rangle & = &  
  \sum_{n=2}^{\infty} \, \sum_{h, \lambda_2,\ldots,\lambda_{n-1},h'} \,
2 \sqrt{\pi N_c} \left(\frac{P^+}{N_c}\right)^{n/2}
\int_{0}^{P^+} dx_{1} \cdots dx_{n}\, 
         \delta\!\left(1 -\sum_{a=1}^{n} x_a\right)  
\nonumber \\
&& \times \ \psi_n(x_1, \ldots, x_n; h, \lambda_2, \ldots \lambda_{n-1}, h') \,
            \nonumber \\
 & &  \times b_{h}^{\dagger}(x_{1}P^+)\,
 a^{\dagger}_{\lambda_2}(x_{2}P^+) \cdots
 a^{\dagger}_{\lambda_{n-1}}(x_{n-1}P^+) \,
d_{h'}^*(x_n P^+)\,  |0\rangle  \ .
\label{Fock}
\end{eqnarray}
The state is  normalised by
\begin{equation}
1   =    \sum_{n=2}^{\infty}\, \sum_{h,\lambda_2,\ldots,\lambda_{n-1},h'}\,
 \int_{0}^{1} dx_{1} \cdots dx_{n} \,
       \delta\!\left(1 -\sum_{a=1}^{n} x_a\right) 
\left| \psi_n \right|^2  \; 
\label{normed}
\end{equation}
if
\begin{equation}
\langle \psi(P^+_1) | \psi(P^+_2)\rangle = 2 P^+_1 (2 \pi)\, 
\delta (P^+_1-P^+_2) \; .
\end{equation}
The object of this paper is to determine the behaviour of the
wavefunctions $\psi_n(x_1,\ldots)$ when one or more
$x_a$ are small. The Fock space matrix elements of the invariant mass
operator $2 P^+ P^-$ resulting from (\ref{ham}) 
were given in Ref.~\cite{mesons}.
For ease of reference and to correct some typing errors in 
interaction {\em (ii)}, we reproduce these in a slightly more
efficient notation as Fig.~\ref{vert}
and Table~\ref{table2} in Appendix~\ref{appendix}.
Projecting 
$2 P^+ P^- |\psi(P^+) \rangle ={\cal M}^2 |\psi(P^+) \rangle  $,
where ${\cal M}$ is the invariant mass eigenvalue,
 onto individual Fock states, one obtains
integral equations for the wavefunctions.

\section{One-link approximation}
\label{onelink}

To begin analysing the small $x$ behaviour of the boundstate equations,
we make the approximation of restricting the sum in Eq.~(\ref{Fock}) to
$n\leq 3$, i.e. at most one flux link in a state.
We correspondingly restrict to  interactions $(i)(ii)(iii)(vii)(viii)$
(see Appendix~\ref{appendix}).
The $\pi$ or $\eta'$ meson state can then be written as
\begin{eqnarray}
  | \psi(P^+)\rangle &=& 2 \sqrt\frac{\pi}{N_c} P^+
\int_0^1 dx\, \psi(x)\,
                 b^\da(x P^+)\, \sigma^2 d^\da((1-x)P^+)\,|0\rangle\nonumber\\
      & & + \frac{2 \sqrt{\pi} (P^+)^{3/2}}{N_c}
\int_0^1 dx \int_0^{1-x}dy \,\sum_\lambda
           \left\{\psi_S(x,y)\, b^\da(xP^+)\, \sigma^2 \, a^\da_\lambda(yP^+)\,
              d^\da((1-x-y)P^+) \right. \nonumber \\
      & & \left. + \psi_A(x,y)\, b^\da(xP^+) \, \zeta^\lambda \sigma^2 \,
             a^\da_\lambda(yP^+) \, d^\da((1-x-y)P^+) \right\}|0\rangle \ .
\end{eqnarray}
To obtain the ${\cal J}_3=0$ spin projection 
component of the $\rho$ meson, one
replaces $\sigma^2 \to \sigma^1$ 
in the above.  
%
%
The wavefunctions of the  ${\cal J}_3=\pm 1$ components of the $\rho$ meson
are more complicated:
\begin{eqnarray}
  | \psi(P^+)\rangle &=& 2 \sqrt\frac{\pi}{N_c} P^+ \int_0^1 dx\, \psi(x)\,
                b^\da(x P^+)\, \sigma^3 d^\da((1-x)P^+)\,|0\rangle\nonumber\\
      & & + \frac{2 \sqrt{\pi} (P^+)^{3/2}}{N_c}
 \int_0^1 dx \int_0^{1-x}dy \,\sum_\lambda
           \left\{\psi_S(x,y)\, b^\da(xP^+)\, \sigma^3 \, a^\da_\lambda(yP^+)\,
              d^\da((1-x-y)P^+) \right. \nonumber \\
      & &  + \psi_A(x,y)\, b^\da(xP^+) \, \zeta^\lambda \sigma^3 \,
             a^\da_\lambda(yP^+) \, d^\da((1-x-y)P^+) \nonumber\\
      & & \left. + \psi_A^\prime(x,y)\, b^\da(xP^+) \, \sigma^1 \zeta^\lambda 
         \sigma^1 \sigma^3 \,
             a^\da_\lambda(yP^+) \, d^\da((1-x-y)P^+) \right\}|0\rangle \; .
\end{eqnarray}
This is the state that is odd under reflections ${\cal P}^1\!: \; x^1\to -x^1$;
the even state is obtained by removing the $\sigma^3$.

The bound state equations for these four states are almost identical.
The $n=2$ bound state equation is
\begin{eqnarray}
 \mathcal{M}^2 \psi(x) &=& \frac{\mu_f^2}{x (1-x)}\, \psi(x) +
   \frac{(\kappa_s^2+\kappa_a^2) N_c}{\pi} \left[
    \int_0^x \frac{dy}{xy} +   \int_0^{1-x} \frac{dy}{(1-x)y}\right] \psi(x)
  \nonumber \\
  & & - \frac{G^2N_c}{2 \pi} \int_0^1 \frac{dy}{(x-y)^2} 
              \left(\psi(y)-\psi(x)\right)  
  \nonumber \\
  & & - \frac{\mu_f \kappa_s \sqrt{N_c}}{\sqrt{\pi}} \left[
        \int_0^x dy\, \frac{(2x-y)\,\psi_S(x-y,y)}{\sqrt{y}\, x (x-y)}
        + \int_0^{1-x} dy\, \frac{(2-2x-y)\,\psi_S(x,y)}{
                            \sqrt{y} (1-x) (1-x-y)}
        \right]
  \nonumber \\
  & & + \frac{\mu_f \kappa_a \sqrt{N_c}}{\sqrt{\pi}} \left[
        \int_0^x dy\, \frac{\sqrt{y}\,\psi_A(x-y,y)}{x (x-y)}
      \right. \nonumber \\ 
  & & \left.  + \int_0^{1-x} dy\, \frac{\sqrt{y}}{(1-x) (1-x-y)}
             \left\{\begin{array}{cl}
                    \psi_A(x,y) & \mbox{for pion} \\
                    -\psi_A(x,y) & \mbox{for rho, ${\cal J}_3=0$} \\
                    -\psi_A^\prime(x,y) & \mbox{for rho, ${\cal J}_3=\pm1$}
          \end{array}\right.
        \right] \; . \label{two}
\end{eqnarray}
For each of the wavefunctions, $\psi_S$, $\psi_A$, $\psi_A^\prime$, the  
$n=3$ bound state equation is
\begin{eqnarray}
 \mathcal{M}^2 \psi(x,y) &=& \mu_f^2\left(\frac{1}{x}+
        \frac{1}{1-x-y}\right)\, \psi(x,y) + \frac{\mu_b^2}{y}\, \psi(x,y) 
  \nonumber \\
  & &  - \frac{G^2N_c}{4 \pi} \left[
               \int_0^{x+y} dz\,\frac{y+z}{\sqrt{yz}(y-z)^2} \, \psi(x+y-z,z)
              \right. \nonumber \\
  & &\hspace{.5in}+ \int_0^{1-x} dz\,\frac{y+z}{\sqrt{yz}(y-z)^2} \, \psi(x,z)
               \nonumber \\
  & &\hspace{.5in}\left.  -2 \psi(x,y)\left(\frac{2}{y}+
                     \int_0^{x+y}\frac{dz}{(y-z)^2} + 
                   \int_0^{1-x}\frac{dz}{(y-z)^2}\right) 
   \right] \;.   \label{three}
\end{eqnarray}
For $\psi_S(x,y)$, we include
\begin{eqnarray}
  & & -\frac{\mu_f \kappa_s \sqrt{N_c}}{\sqrt{\pi y}}\left[
        \frac{2x+y}{x(x+y)}\, \psi(x+y)+ \frac{2-2x-y}{(1-x)(1-x-y)}\, \psi(x)
        \right] \; .
\end{eqnarray}
For $\psi_A(x,y)$ in the pion and the rho ${\cal J}_3=0$ state, we include
\begin{eqnarray}
  & & +\frac{\mu_f \kappa_a \sqrt{N_c y}}{\sqrt{\pi}}\left[
        \frac{\psi(x+y)}{x(x+y)}+ \frac{1}{(1-x)(1-x-y)}\,
            \left\{\begin{array}{cl}
                    \psi(x) & \mbox{for pion}\\
                    -\psi(x) & \mbox{for rho}
                  \end{array}\right.
        \right] \; .
\end{eqnarray}
For $\psi_A(x,y)$ in the rho ${\cal J}_3=\pm1$ states, we include
\begin{eqnarray}
  & & +\frac{\mu_f \kappa_a \sqrt{N_c y} \, \psi(x+y)}{\sqrt{\pi} x (x+y)}\; .
\end{eqnarray}
For $\psi_A^\prime(x,y)$ in the rho ${\cal J}_3=\pm1$ states, we include
\begin{eqnarray}
 & & -\frac{\mu_f \kappa_a \sqrt{N_c y} \, \psi(x)}{\sqrt{\pi} (1-x) (1-x-y)}\; .
\end{eqnarray}

\begin{figure}
\input{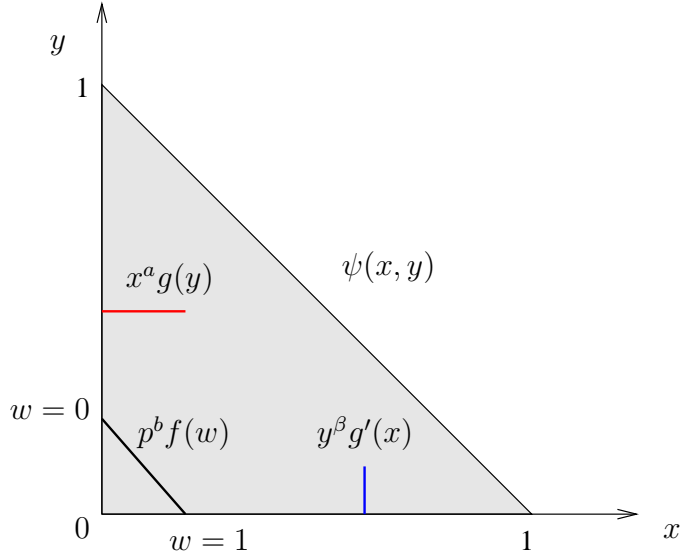}
\caption{The $n=3$ wavefunction $\psi(x,y)$ in
the limits $x\to 0$, $y \to 0$, or $p=x+y \to 0$. \label{psixy}}
\end{figure}

Next, we analyze the singularities  as one or more
of the momenta go to zero.  We will consider each of the following
limits:
\begin{eqnarray}
    \lim_{x\to 0} \psi(x) &=& C_2 \, x^\alpha  \\
    \lim_{\stackrel{x\to 0,}{ y \;\mathrm{constant}}} 
            \psi(x,y) &=& x^{a} g(y) \\
    \lim_{\stackrel{y\to 0,}{x \;\mathrm{constant}}} 
          \psi(x,y) &=& y^\beta g^\prime\!(x)\label{yto0An}\\
    \lim_{\stackrel{x+y\to 0,}{x/y \;\mathrm{ constant}}} 
        \psi(x,y) &=& (x+y)^b \, f\!\left(\frac{x}{x+y}\right) = p^b f\!(w)
       \; , \label{xylimit}
\end{eqnarray}
where $x = wp$, and $y = (1-w)p$.  See Figure~\ref{psixy}.
The remaining limit of interest, $1-x\to 0$ with $(1-x)/y$ constant, 
is equivalent to the limit (\ref{xylimit}) by charge conjugation.

\subsection{$n=3$ equation, limit $x\to 0$:}
\label{threexzero}
For $\psi_S(x,y)$, we obtain:
\begin{eqnarray}
	0&=&\mu_f^2 \,\,g_S\!(y)\,x^{a-1} - \frac{\mu_f\kappa_s\sqrt{N_c}}{\sqrt{\pi y}}\psi(y) x^{-1}\nonumber\\
&&-\frac{G^2N_c}{4\pi}\int^{x+y}_0 \frac{\mathrm{d}z}{(y-z)^2}\left[\frac{y+z}{\sqrt{yz}}\psi(x+y-z)-2\psi(x,y)\right]\nonumber\\
&&+O(x^a)\; .
\end{eqnarray}
For $\psi_A(x,y)$, we obtain:
\begin{eqnarray}
	0&=&\mu_f^2 \,\,g_A\!(y)\,x^{a-1} +\frac{\mu_f\kappa_a\sqrt{N_c}}{\sqrt{\pi y}}\psi(y) x^{-1}\nonumber\\
&&-\frac{G^2N_c}{4\pi}\int^{x+y}_0 \frac{\mathrm{d}z}{(y-z)^2}\left[\frac{y+z}{\sqrt{yz}}\psi(x+y-z)-2\psi(x,y)\right]\nonumber\\
&&+O(x^a)\; .
\end{eqnarray}
In order to cancel the divergence of second term, we must set $a=0$. 
We then find that the integral term, which is also $O(x^{a-1})$, goes to zero.
Simplifying, we get:
\begin{eqnarray}
\lim_{x \to 0} \psi_S(x,y)&=& \frac{\kappa_s\sqrt{N_c}}{\mu_f\sqrt{\pi y}}\psi(y) \label{fzms}\\
 \lim_{x \to 0} \psi_A(x,y)&=& -\frac{\kappa_a\sqrt{N_c}}{\mu_f\sqrt{\pi y}}\psi(y) \; . \label{fzma}
\end{eqnarray}

\subsection{$n=3$ equation, limit $y\to 0$:}
\label{threeyzero}
For this limit, the resulting equation is the same for $\psi_S$ and $\psi_A$ 
because the differing terms are non-leading.  
\begin{eqnarray}
	0&=&\left(\mu_b^2+\frac{G^2N_c}{\pi}\right)\,g^\prime\!(x)\,y^{\beta-1}\nonumber\\
&&-\frac{G^2N_c}{4\pi}y^{-1}\int^{1/y}_0 \frac{\mathrm{d}q}{(1-q)^2}\left[\frac{1+q}{\sqrt{q}}(qy)^\beta g^\prime(x) -2 y^\beta g^\prime(x)\right]\nonumber\\
&&+O(y^{-\frac{1}{2}})\mbox{.} 
\end{eqnarray}
We find the constraint $\beta<\frac{1}{2}$, so that the integrals 
converge, in which case the $O(y^{-\frac{1}{2}})$ terms are non-leading.   
We also have the constraint $\beta>0$, so that the expectation value
of $2P^+P^-$  is finite.
Thus, we obtain:
\begin{eqnarray}
  0&=&\left(\mu_b^2+\frac{G^2N_c}{\pi}\right)\,g^\prime\!(x)\,
              y^{\beta-1}
  -\frac{G^2N_c}{4\pi}\,g^\prime\!(x)\,y^{\beta-1}\,\,
     2\left[2\pi\beta\tan{(\pi\beta)}+2\right] \; .
\end{eqnarray}
Simplifying, we derive:
\begin{equation}
0=\left[\mu_b^2 - G^2N_c\beta\tan{(\pi\beta)}\right]\,
   g^\prime\!(x)\,y^{\beta-1}\mbox{.}\label{twogluon}
\end{equation}
Note that $g^\prime\!(x)$ is not fixed by this analysis.  
Eq.~(\ref{twogluon}) is the same expression that one
obtains for the two-link  bound state equation without quarks~\cite{bardeen}.

\subsection{$n=3$ equation, limit  $x+y\to 0$, 
             $x/y\;\mathrm{constant}$}
\label{threexyzero}
In this limit, we obtain, for both $f_S(w)$ and $f_A(w)$:
\begin{eqnarray}
   0&=&\frac{\mu^2_f\,f(w)}{w} p^{b-1} + 
      \frac{(\mu^2_b+\frac{G^2N_c}{\pi})\,f(w)}{1-w}p^{b-1}\nonumber\\
   &&-\frac{G^2N_c}{4\pi}\, p^{b-1} \int^1_0\frac{\mathrm{d}q}{(q-w)^2}
      \left[\frac{(2-w-q)\,f(q)}{\sqrt{(1-w)(1-q)}}-2\,f(w)\right]
      \nonumber\\
   &&-\frac{G^2 N_c}{4\pi}\, p^{b-1}\int^1_0\frac{\mathrm{d}q}{(q-w)^2}
      \left[\frac{(w+q-2 q w)\,f(q)}{\sqrt{(1-w)(1-q)}}
           \left(\frac{w}{q}\right)^\alpha -2w\,f(w)\right]
      \nonumber\\
   && + \frac{\mu_f C_2 \sqrt{N_c} }{\sqrt{\pi} w}\,p^{\alpha-3/2} \left\{\begin{array}{cl}
         -\kappa_s \frac{(w+1)}{\sqrt{1-w}} 
            & \mbox{for $f_S(w)$ }\\
               \kappa_a \sqrt{1-w} 
            & \mbox{for $f_A(w)$} 
        \end{array}\right. \nonumber\\
    && + O\left(p^{-1/2}\right)  \; . \label{fw}
\end{eqnarray}
In order for the last term in the equation to be cancelled in the limit 
$p\to0$, we must set $b=\alpha-\frac{1}{2}$. Thus,
\begin{equation}
  \lim_{x+y\to 0} \psi(x,y)= p^{\alpha-1/2} f(w) \; . \label{alpha}
\end{equation}

\subsection{$n=2$ equation, limit ${x\to 0}$:}
\label{twoxzero}
Taking the $x\to 0$ limit of Eq.~(\ref{two}) and using (\ref{alpha}), 
we obtain
\begin{eqnarray}
  0&=&\mu_f^2 C_2 x^{\alpha-1} - \frac{G^2N_c C_2}{2\pi}
       \left(1-\pi\alpha\cot{(\pi\alpha)}\right) x^{\alpha-1}\nonumber\\
  &&+x^{\alpha-1} \int^1_0 \frac{\mathrm{d}z}{1-z}\left[
        \frac{\kappa_s^2 N_c C_2}{\pi}-\frac{\mu_f \kappa_s \sqrt{N_c}}{
         \sqrt{\pi}}\left(\frac{(2-z)\,f_{\!S}(1-z)}{\sqrt{z}}
                                        \right)\right]\nonumber\\
   &&+x^{\alpha-1} \int^1_0 \frac{\mathrm{d}z}{1-z}\left[\frac{
           \kappa_a^2 N_c C_2}{\pi}+\frac{\mu_f \kappa_a \sqrt{N_c}}{
             \sqrt{\pi}}(\sqrt{z}\,f_{\!A}\!(1-z))\right]\nonumber\\
&&+O(x^0) \; . \label{twoeqn}
\end{eqnarray}
For the expectation value of the energy to be finite, we must have 
$0 \le \alpha <1$.

\subsection{Self-consistent solution}

\begin{figure}
\begin{center}
\begin{tabular}{c@{}c}
$ \displaystyle \frac{f_S(w)\, \mu_f}{C_2\, \kappa_s \sqrt{N_c}}$ &
\BoxedEPSF{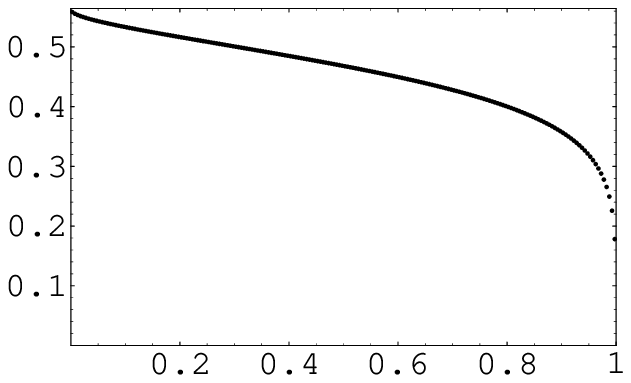}\\
 &  $w$ \end{tabular} \\
\begin{tabular}{c@{}c}
$ \displaystyle \frac{f_A(w)\, \mu_f}{C_2\, \kappa_a \sqrt{N_c} }$ &
\BoxedEPSF{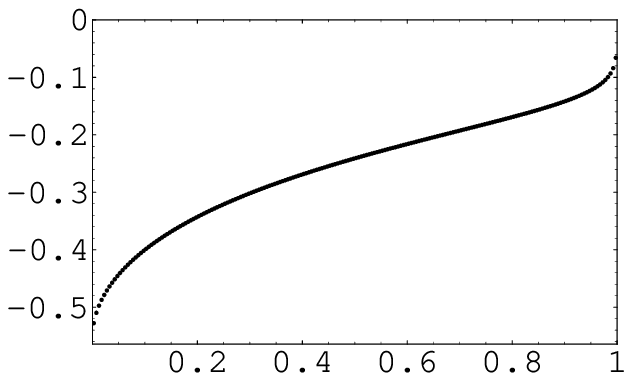}\\  & $w$
\end{tabular}\end{center}

\caption{$f_S(w)$ and $f_A(w)$ vs.\ $w$ for couplings 
$\mu_f=0.362 G \sqrt{N_c}$, $\mu_b= 0.2 G \sqrt{N_c}$, and $\alpha=0.4$.}
\end{figure}

\begin{figure}
\begin{center}
$\alpha$\BoxedEPSF{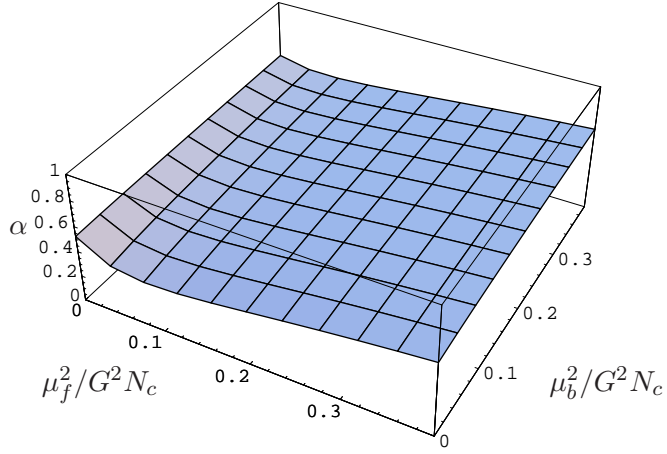 scaled 750} \\[-0.5in]
\mbox{}\hspace{0.5in}$\mu_f^2/G^2 N_c$ \hspace{2in}$\mu_b^2/G^2 N_c$
\end{center}
\caption{$\alpha$ vs. $\mu_f^2$ and $\mu_b^2$,  $\kappa_s 
=-0.323 G $, $\kappa_a = 0.162 G $. \label{figa}}
\end{figure}

\begin{figure}
\begin{center}
$\alpha$\BoxedEPSF{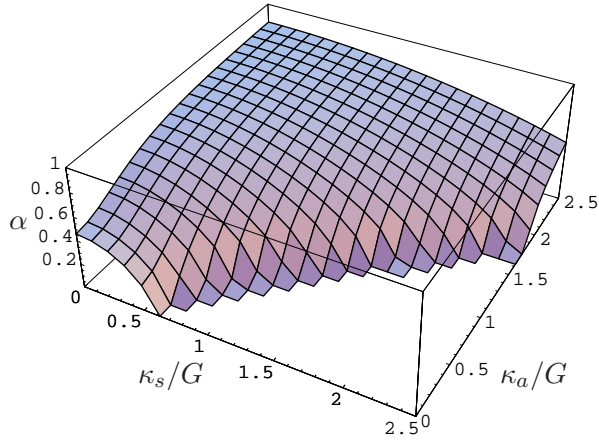 scaled 750}\\[-0.5in]
\mbox{}\hspace{0.5in}$\kappa_s /G$ \hspace{1.5in}$\kappa_a /G$
\end{center}
\caption{$\alpha$ vs.\ $\kappa_s /G$ and $\kappa_a /G$;  
$\mu_f = 0.362 G 
\sqrt{N_c}$, and $\mu_b = 0.2 G \sqrt{N_c}$.
\label{figb}}
\end{figure}

To determine $\alpha$, one must
find $f_S(w)/C_2$ and $f_A(w)/C_2$ that self-consistently
solve Equations (\ref{fw}) and (\ref{twoeqn}).
We obtained solutions numerically by discretizing $w$ on the interval [0,1]. 
Using numerical calculations of $f_S(w)/C_2$ and $f_A(w)/C_2$, one can study
the dependence of $\alpha$ on the couplings in the Hamiltonian; 
see Figures~\ref{figa}--\ref{figb}.
Note that $\alpha$ is very nearly constant with
respect to link field mass $\mu_b$ (Fig.~\ref{figa}).
Also, note that $\alpha$ is outside of its allowed range
for some areas of the coupling space,
corresponding generally to large values of $\kappa_s \sqrt{N_c}$ and
$\kappa_a \sqrt{N_c}$ (Fig.~\ref{figb}).  Thus, the endpoint behavior
of the wavefunction can restrict the allowed values of the couplings.

\section{Beyond one link}
\label{analytic}

When relaxing the truncation on $n$ and including the other
interactions of Fig.~\ref{vert}, most of the complication
occurs in the
analytic auxiliary
functions, such as $f$, $g$, $g^\prime$ in the previous section. If
we wish to determine only the form of the 
leading non-analytic behaviour of wavefunctions
at small momenta, the analysis is somewhat simpler. We note that
this is independent of spin, represented by quark helicity and 
link field transverse 
orientation.
Consider the effect of including $n=4$ wavefunctions $\psi_4$
and the other
interactions of Fig.~\ref{vert}. 
In the following we ignore the spin dependence, 
finding only the 
momentum dependent part
of these contributions. Accordingly, we label wavefunction 
components $\psi_n$ by the number of partons $n$ and their lightcone
momenta only.

\subsection{$n=2$ equation}
To the $n=2$  boundstate equation (\ref{two}) we must add interactions
$(iv)(vi)$:
\begin{equation}
 \int dy  dz \ \frac{y-z}{ 4 \pi (y+z)^2 \sqrt{y z}} 
\psi_4(x-y-z,y,z) + {\rm antiquarks}  
\end{equation}
from $(iv)$, and
\begin{equation}
 \int dy dz \ \frac{1}{ 4 \pi (x-z)\sqrt{yz}} 
\psi_4(x-y-z,y,z) +  {\rm antiquarks}  
\end{equation}
from $(vi)$.
If we assume that these contribute at the leading $x^{\alpha-1}$
order as $x \to 0$ in the analsyis of Section~\ref{twoxzero},
then 
\begin{equation}
\psi_4(x,y,z) \sim (1-x)^{\alpha -1} \ {\rm as} \ x \to 1 \ .
\label{3zero}
\end{equation} 

\subsection{$n=3$ equation}
In the $n=3$ equation (\ref{three}), we must include interaction $(v)$, 
that connects
$n=3$ to $n=3$ and was previously omitted:
\begin{equation}
 \int dz \ \frac{1}{ 4 \pi (x+y)\sqrt{yz}} 
\psi_3(x+y-z,z) +  {\rm antiquarks}  \ .
\end{equation}
This is subleading in the $x \to 0$ analysis of Section~\ref{threexzero} and
the $y \to 0$ analysis of Section~\ref{threeyzero} (recall that $\beta <1/2$).
It contributes at the leading order $p^{b-1}$ in the analysis 
of Section~\ref{threexyzero}, but does not affect the relation
$b=\alpha-\frac{1}{2}$.
To the  $n=3$ equation (\ref{three}) we must also add 
terms involving $\psi_4$ via interaction $(ii)$:
\begin{equation}
 \int dz \ \frac{1 }{ 2 \sqrt{\pi z}} 
\left(\frac{1}{ x} \pm \frac{1}{ x-z}\right)
\psi_4(x-z,z,y) \ .
\label{ii}
\end{equation}
If we assume that these contribute at the leading $x^{-1}$
order of the analysis in Section~\ref{threexzero},
then 
\begin{equation}
\psi_4(wp,(1-w)p,z) \sim p^{-1/2} \ {\rm as} \ p \to 0 \ .
\label{asump}
\end{equation} 
In the $y \to 0$ analysis of Section~\ref{threeyzero}, 
(\ref{ii}) does not contribute
at leading order since, by normalisability, $\beta-1<-1/2$ and
$\psi_4(x-z,z,y)$ cannot
diverge faster than $1/\sqrt{y}$ when $y \to 0$.
In the analysis 
of Section~\ref{threexyzero}, property (\ref{3zero}) implies that (\ref{ii})
contributes at the leading $p^{\alpha-3/2}$
order as $p \to 0$.

\subsection{$n=4$ equation}
Finally, there will be a new $n=4$ boundstate equation 
for ${\cal M}^2 \psi_4 (x,y,z)$.
The momentum dependent parts of these contributions are given in 
Table~\ref{table1}.
\begin{table}
\caption{The momentum dependent parts of interactions 
contributing to the boundstate equation for $\mathcal{M}^2 \psi_4(x,y,z)$.
\label{table1}}
\centering$\displaystyle
\renewcommand{\arraystretch}{1.25}
\begin{array}{|c|c|}
\hline
\frac{1}{ x} \psi_4(x,y,z) 
+  {\rm antiquarks}  & (i)  \\
\hline
  \frac{1}{ 2 \sqrt{\pi y}} \left(\frac{1}{ x} \pm \frac{1}{ x+y}\right)
\psi_3(x+y,z) +  {\rm antiquarks}  
& (ii) \\
\hline
  \frac{(y-z)\psi_2(x+y+z) 
  }{ 4 \pi (y+z)^2 \sqrt{y z}} 
+ {\rm antiquarks}  & (iv) \\
\hline
 \int dv \ \frac{\psi_4(x+y-v,v,z) 
 }{ 4 \pi (x+y)\sqrt{yv}} 
+  {\rm antiquarks}  & (v) \\
\hline
  \frac{\psi_2(x+y+z) }{ 4 \pi (x+y)\sqrt{yz}} 
+  {\rm antiquarks}  & (vi) \\
\hline
\left(\frac{1}{ y} + \frac{1}{ z}\right) 
\psi_4(x,y,z) 
& (vii) \\
\hline
\frac{-1 }{ 4 \pi} {\rm P}\int dv \  \frac{(y+v) 
\psi_4(x+y-v,v,z)}{ \sqrt{v y} (v-y)^2}
+  {\rm antiquarks} & (viii) \\
\hline
\frac{-1 }{ 8 \pi} {\rm P}\int dv \ \frac{(v+z)(2y+z-v) 
\psi_4(x,y+z-v,v)}{\sqrt{yzv(y+z-v)} 
(z-v)^2} 
& (ix) \\
\hline
\int dv \ \frac{\psi_4(x,y+z-v,v)  
}{ 4 \pi \sqrt{yzv(y+z-v)}} 
& (x) \\
\hline
\frac{-1 }{ 8 \pi} \int dv \ \frac{(y-z)(y+z-2v) }{ \sqrt{yzv(y+z-v)}  (z+y)^2}
\psi_4(x,y+z-v,v)  
& (xiii) \\
\hline
\end{array}$
\end{table}

Repeating the analysis of Sections~\ref{threexzero} and 
\ref{threeyzero} for $n=4$, we again find 
from comparing interactions $(i)$ and $(ii)$ that
\begin{equation}
\psi_4(x,y,z)  \sim x^{0}  \ {\rm as} \ x \to 0 \ ,
\end{equation} 
since all other interactions are 
subleading, while from interactions $(viii)(ix)$ 
\begin{equation}
\psi_4(x,y,z)  \sim y^{\beta}  \ {\rm as} \ y \to 0 \ .
\end{equation} 
We confirm directly, from the $n=4$ amplitudes of Table I,
the correctness of Eq.~(\ref{asump}) and Eq.~(\ref{3zero}), thus 
verifying the  assumptions that went into deriving them. In a similar
way we also deduce as $p\to 0$:
\begin{eqnarray}
&\psi_4(x,wp,(1-w)p)  & \sim p^{0}  \\
&\psi_4(wp,y,(1-w)p)  &\sim p^{\beta}  \\
&\psi_4(wp,y,1-y-p)  &\sim p^{0}
\end{eqnarray}
and 
\begin{equation}
\psi_4(x,y,z)  \sim (1-y)^{\alpha-1/2} \ {\rm as} \ y \to 1
\end{equation} 

\subsection{Summary}
We can summarize all these results in 
the following forms for $\psi_n$, which 
have the correct endpoint behaviour when any number of momenta vanish:
\begin{eqnarray}
\psi_2(x) & = & C_2 x^{\alpha} (1-x)^{\alpha} \nonumber \\
\psi_3(x,y) & = & C_3 \frac{y^{\beta} (1-y)^{\alpha}}{
[(x+y)(1-x)]^{\beta-\alpha+1/2}} \label{ansatz}
\\
\psi_{4}(x,y,z) & =&   C_4  \frac{(yz)^{\beta} [(1-y)(1-z)]^{\alpha} 
[(x+y+z)(1-x)]^{\alpha+\beta-1/2}}{  [(x+ y)(1-x-y)]^{\beta  +1/2}(y+z)^{2\beta}  }  \nonumber 
\end{eqnarray}
The exponents $\alpha$ and $\beta$ are independent of $n$, quark helicity,
and the transverse shape, while
the auxiliary functions $C_n$ are dependent on all of these factors. 
If the $C_n$ are made from a complete set of functions of $x_a$, non-singular
in the limit when any number of momenta vanish, then any boundstate
wavefunction should be expressible in the form (\ref{ansatz}).

We have also 
examined the effect of higher orders in $n\geq 5$. 
The forms (\ref{ansatz}) remain invariant, although the exponents $\alpha$
and $\beta$ are renormlised in general because  of parton self-energy and 
vacuum polarization.
Wavefunctions
$\psi_n$ analogous to (\ref{ansatz}) become increasing complicated at higher 
 $n$
and we have not found a simple general expression; no new exponents beyond
$\alpha$ and $\beta$ occur however.
Of special interest are the limits
\begin{eqnarray}
& \psi_n(x_1,x_2, \ldots)   \sim (1-x_1)^{\alpha-(n-2)/2} & 
{\rm as} \ x_1 \to 1
\label{uno}\\
& \psi_n(x_1,x_2, \ldots)   \sim (x_1)^{0} & {\rm as} \ x_1 \to 0
\label{nought}
\end{eqnarray}
which follow from comparing interactions $(i)$ and  $(ii)$. These
govern the behaviour of hadronic structure functions at large
and small Bjorken $x$, as discussed in Section~\ref{appl}.

The forms (\ref{ansatz}) demonstrate that
lightcone wavefunctions become more singular as more
lightcone momenta vanish, even though this is a `high-energy' limit. 
Intuition from
the free lightcone kinetic energy (sum of all terms $(i)$ and $(vii)$ 
in Table~\ref{table2} of Appendix~\ref{appendix})
would have suggested the complete opposite. Rather than simply vanishing
at the edges of phase space, as indicated by the free theory,
the interactions have forced 
relations between wavefunctions with different numbers of partons to cancel
divergences in the boundstate equations. This was noted for continuum QCD
for a single vanishing $x_a$ in Ref.~\cite{ladder},
but for the transverse lattice we are able to generalise it to any
number of vanishing momenta. Our analysis rules out a number of naive
models for the lightcone wavefunctions, including products of individual
parton wavefunctions and functions of the free kinetic energy.

\section{Applications}
\label{appl}

Based on the previous analysis, we can make
definite statements about the behaviour of
quark parton distributions in mesons at the edges of phase space.
Some other recent studies of these distributions in various
models have been made in Refs.~\cite{other}.
We use the generalised quark 
distributions \cite{diehl}, which in lightcone gauge are defined as
\begin{equation}
{\cal H}(\overline{x},\xi,Q^2)=
\frac{1}{ \sqrt{1-\xi^2}} \int_{-\infty}^{+\infty} \frac{dz^-}{ 4 \pi}
{\rm e}^{{\rm i} \overline{x} \overline{P}^{+} z^-}
\langle P_{\rm in} | \overline{\Psi}(-z^-/2) \gamma^+
\Psi({z}^-/2) | P_{\rm out} \rangle \ ,
\label{gpd}
\end{equation}
\begin{eqnarray}
Q & = &  P_{\rm in} - P_{\rm out} \\ 
\xi & = & \frac{( P_{\rm in} - P_{\rm out})^+ }{ 2\overline{P}^+} \\
\overline{P}^+ & = & \frac{(P_{\rm in}  + P_{\rm out})^+ }{ 2} \ .
\end{eqnarray}
Bars generally indicate an average over in and out states.
In Eq.~(\ref{gpd}), we define the boundstates at non-zero transverse
momentum ${\bf P}$ by boosting the Fock states
with the Poincar\'e generators ${\bf M}^{+} = (M^{+1}, M^{+2})$: 
\begin{eqnarray} 
&& {\rm exp}\left[ -{\rm i} {\bf M}^{+}. {\bf P}/P^+ \right]
b_{h}^{\dagger}(x_{1}P^+)
a^{\dagger}_{\lambda_2}(x_{2}P^+) \cdots
 a^{\dagger}_{\lambda_{n-1}}(x_{n-1}P^+)
d_{h'}^*(x_n P^+)  |0\rangle   \nonumber \\
& = &
{\rm exp}\left[ {\rm i} {\bf P}. {\bf c} \right]
b_{h}^{\dagger}(x_{1}P^+)
a^{\dagger}_{\lambda_2}(x_{2}P^+) \cdots
 a^{\dagger}_{\lambda_{n-1}}(x_{n-1}P^+)
d_{h'}^*(x_n P^+)  |0\rangle \ ,  
\label{boost}
\end{eqnarray}
\begin{eqnarray}
{\bf c} = (c^1 , c^2) & = & \sum_{a=1}^{n} x_a {\bf x}_{a} \ , \\
{\bf x}_{p+1} & = &  {\bf x}_{p} + a \hat{\lambda}_{p+1} \ , \ p=1 \ {\rm to} \
n-2 \ ,
\end{eqnarray}
where $\hat{\lambda}_{p+1}$ is a transverse 
unit vector in the direction of the $p^{\rm th}$
link. The boundaries of interest are typically at large and small 
$\overline{x}$ and at the transition point $\overline{x} = \xi$.

\subsection{$\overline{x} \to 1$}
We find from Eqs.~(\ref{gpd}) and (\ref{uno}) that
each $\psi_n$ contributes $\sim  (1-\overline{x})^{2 \alpha}$ to
${\cal H}(\overline{x},\xi, Q^2)$ as $\overline{x} \to 1$. 
This is interesting because the conventional wisdom, based upon the
free kinetic energy, is that 
$\psi_2$ should dominate in this region --- all but one parton has
small momentum, so the free kinetic energy is minimized by the minimum
number of partons. Although the exponent {\em is}\/ correctly given by
the lowest Fock wavefunction, which governs the distribution amplitude,
on the transverse lattice all higher
Fock states contribute to the structure function with the same power. 

\subsection{$x \to 0$}
We find from Eqs.~(\ref{gpd}) and (\ref{nought}) that
$\psi_n$ contributes $\sim (\ln \frac{1 }{ x})^{n-3}$, $n > 2$,  
to the traditional quark distribution
function
${\cal H}({x},0,0)$ as ${x} \to 0$. Thus, the sum over all Fock sectors
$n$ rises faster than any power of $\ln x$, i.e. at least like a power
of $x$. This is consistent with the Regge behaviour of structure 
functions in the small-$x$ region.

\subsection{$\overline{x} \to \xi^+$}
Unfortunately, we are not able to explore the region $\overline{x} < \xi$
with the strict large-$N_c$ wavefunctions. That region is governed by
their $1/N_c$ corrections, since it depends upon quark pair production.
However, we find from Eq.~(\ref{ansatz}) that
$\psi_2$ contributes zero, while 
$\psi_3$ and $\psi_4$ contribute  finite constants to 
${\cal H}(\overline{x},\xi, Q^2)$ as $\overline{x} \to \xi$ from above,
provided $\xi > 0$.
This is presumably true for higher $n$ also and,
provided these constants fall sufficiently rapidly with $n$,
suggests that ${\cal H}(\xi,\xi,Q^2)$ is finite.

\section{Conclusions}
\label{last}
By demanding finiteness of the boundstate energy, we have determined the
form (\ref{ansatz})
of meson wavefunctions at small lightcone momentum in a coarse 
transverse lattice
gauge theory at large $N_c$. The forms depend on two exponents
$\alpha$ and $\beta$, independent of the other details
of the wavefunction such as spin and number of partons, 
which encode the leading non-analytic behaviour.
As more momenta vanish,
wavefunctions become more singular, contrary to the naive intuition
from free lightcone kinetic energy, which diverges in these limits. 
The analytic forms lead to predictions for the behaviour near their
endpoints of quark distributions. In particular, we see Regge behaviour
at small $x$ and contributions from all Fock states at large $x$.

The use of coarse lattice wavefunctions in transverse position space simplified
the analysis, allowing us to separate the lightcone momentum 
problem from the transverse structure. The large-$N_c$ limit also
aided us by restricting the diagrammatic rules to planar ones and
suppressing quark pair production.
We would need to reformulate the problem at finite $N_c$ in order to study
the small lightcone momentum behaviour in baryons and in the sea quark 
part of wavefunctions. 
There is also the question of higher orders of the colour-dieletric expansion;
that is, higher powers of dynamical quark and link fields in the
lightcone hamiltonian. We have investigated some of these possible
terms, but found none
that altered the form (\ref{ansatz}), although the values of $\alpha$
and $\beta$ are affected. 
Finally, it would be interesting to investigate similar questions in continuum
lightcone QCD.

\acknowledgments{
This work was supported in part by PPARC Visiting Fellowship
grant PPA/V/S/2000/00513.
The work of SD was supported by PPARC grant PPA/G/0/2002/00470.
The work of JB, BvdS, \& EW was supported by 
the Research Corporation.
}

\newpage

\appendix


\section{Interactions}
\label{appendix}

\begin{figure}[hp]
\centering
\BoxedEPSF{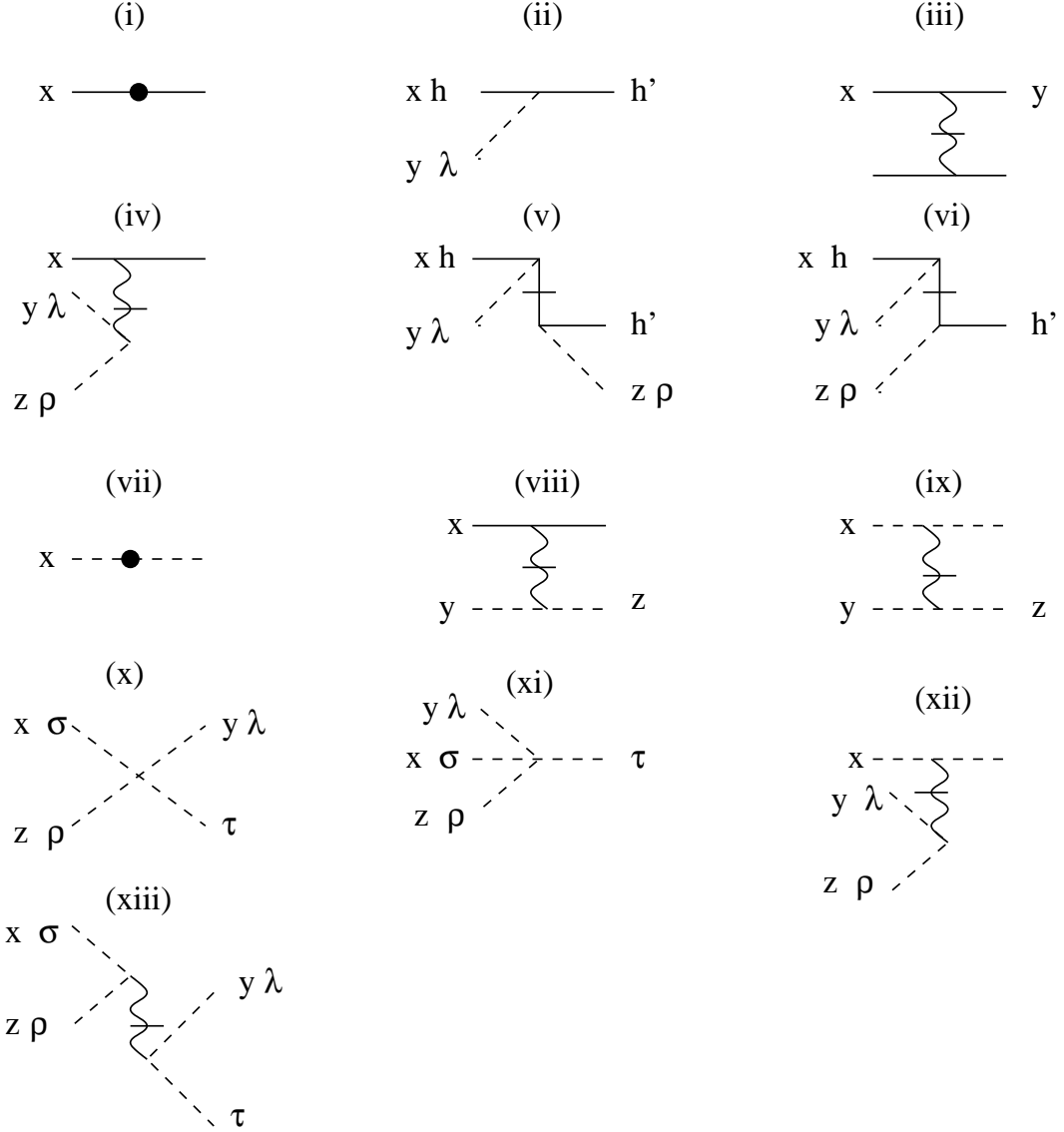 scaled 700}
\caption{ Planar diagram representation of the elementary interaction
vertices following  at large $N_c$.
Time $x^+$ is increasing to the right. 
Solid lines are quarks, chain lines
are link fields. $x,y,z$ represent lightcone momentum fractions, $h,h'$
helicities, $\lambda,\sigma, \rho, \tau \in \{\pm 1, \pm 2\}$ transverse
orientation of links. 
Vertical barred lines are $x^+$-instantaneous
interactions. Spectator partons in the hadron are not drawn.
\label{vert}}
\end{figure}

\newpage

\begin{table}[hp]
\caption{Matrix elements of the invariant mass operator
$2P^+P^-$ (for zero tranverse momentum) in Fock space; see Fig.~\ref{vert} for
corresponding diagrams of interaction vertices. 
Momentum conserving delta functions
are omitted for clarity. P stands for principal part.
$\perpto{\lambda,\rho}=1$ if $|\lambda|\neq |\rho|$ and 0 otherwise.
\label{table2}}
\centering$\displaystyle
\renewcommand{\arraystretch}{1.25}
\begin{array}{|c|c|}
\hline
\frac{1}{x}\left(\mu_{f}^{2}  + \frac{(\kappa_{a}^{2} +
\kappa_{s}^{2}) N_c}{ \pi} \int_{0}^{x} \frac{dy}{ y} \right)& (i)  \\
\hline
\frac{1}{ 2\sqrt{\pi y}}\left\{ -\left(\frac{1}{x+y} +
\frac{1}{ x} \right) \mu_f \kappa_s \sqrt{N_c} \,\delta_{hh'} 
+ \left(\frac{1}{ x+y} -
\frac{1}{ x} \right)  \mu_f \kappa_a \sqrt{N_c} \,
      {\zeta^\lambda}_{h^\prime h}
                                 \right\} & (ii) \\
\hline
\frac{-G^2 N_c }{ 2 \pi} \ {\rm P}\left( \frac{1}{ (x-y)^2} \right) & (iii) \\
\hline
\frac{-G^2 N_c
(y-z)}{ 4 \pi (y+z)^2 \sqrt{y z}} \,\delta_{-\lambda \rho} & (iv) \\
\hline
\frac{N_c}{ 4 \pi (x+y)\sqrt{yz}}\left\{ \kappa_{s}^{2} \delta_{hh'} +
\kappa_{a}^{2} \,
    \left(\zeta^\rho \zeta^\lambda\right)_{h^\prime h}
    + \kappa_a \kappa_s \left(\zeta^\rho + \zeta^\lambda\right)_{h^\prime h}
                                       \right\} & (v)  \\
\hline
{\rm As} \ (v) \ {\rm with} \ \rho \to -\rho & (vi)  \\ 
\hline
\frac{m_{b}^{2}}{ x} & (vii) \\
\hline
\frac{-G^2 N_c}{ 4 \pi} {\rm P}\left( \frac{y+z}{ \sqrt{z y} (z-y)^2}
\right) & (viii) \\
\hline
\frac{-G^2 N_c}{ 8 \pi} {\rm P}\left( \frac{(y+z)(2x+y-z)}{ \sqrt{xzy(x+y-z)} 
(z-y)^2} \right) & (ix) \\
\hline
\frac{1}{ 4 \pi \sqrt{xyz(x+z-y)}} \left\{ 2\lambda_1 \delta_{\sigma \lambda}
\delta_{\rho \tau} \delta_{-\rho \sigma} +  \lambda_2 ( \delta_{\sigma \lambda}
\delta_{\rho \tau} \delta_{\rho \sigma} + \delta_{-\sigma \rho}
\delta_{-\lambda \tau} \delta_{-\sigma \lambda}) \right. & (x) \\ 
 \left. +\lambda_4 (\delta_{\sigma \lambda}
\delta_{\rho \tau}
\perpto{\sigma,\rho}                         
+\delta_{-\sigma \rho}
\delta_{-\lambda \tau}
\perpto{\sigma,\lambda}                      
)
-\beta\delta_{\lambda \rho} \delta_{\sigma \tau}
\perpto{\sigma,\rho}                         
\right\} & \\
\hline
\frac{1}{ 4 \pi \sqrt{xyz(x+z+y)}}
 \left\{ 2\lambda_1 \delta_{-\sigma \lambda}
\delta_{\lambda \tau} \delta_{\lambda \rho } +  \lambda_2  \delta_{\sigma \tau}
\delta_{-\lambda \rho} \delta_{|\lambda| |\sigma|} \right. & (xi) \\
 \left. + \lambda_4 
(\delta_{-\lambda \sigma} \delta_{\rho \tau}
\perpto{\lambda,\rho}                        
+ \delta_{-\sigma \rho}
\delta_{\lambda \tau}
\perpto{\sigma,\lambda}                      
)
-\beta\delta_{-\lambda \rho} \delta_{\sigma \tau}
\perpto{\sigma,\lambda}                      
\right\} & \\
\hline
\frac{-G^2 N_c}{ 8 \pi} \frac{(y-z)(2x+y+z)}{ \sqrt{xzy(x+y+z)} (z+y)^2}
\delta_{-\lambda \rho} & (xii) \\
\hline
\frac{-G^2 N_c}{ 8 \pi} \frac{(x-z)(2y-x-z)}{ \sqrt{xzy(x-y+z)} (z+x)^2}
\delta_{-\sigma \rho}\delta_{-\lambda \tau} & (xiii) \\
\hline
\end{array}$
\end{table}



\begin{thebibliography}{}

\bibitem{bardeen}
W.~A.~Bardeen and R.~B.~Pearson,
Phys.\ Rev.\ D {\bf 14}, 547 (1976).

\bibitem{review} 
M. Burkardt and S. Dalley, 
Prog.\  Part.\ Nucl.\ Phys.\   {\bf 48}, 317 (2002)
[arXiv:hep-ph/0112007].

\bibitem{hoof} 
G. 't Hooft, 
Nucl.\ Phys.\ B {\bf 75}, 461 (1974).

\bibitem{ladder} 
F. Antonuccio, S. J. Brodsky, and S. Dalley,
Phys. Lett. B {\bf 412}, 104 (1997)
[arXiv:hep-ph/9705413].

\bibitem{colour}
S.~Dalley and B.~van de Sande,
Phys.\ Rev.\ D {\bf 56}, 7917 (1997)
[arXiv:hep-ph/9704408].

\bibitem{morris}
S.~Dalley and T. R. Morris,
Int. J. Mod. Phys. A {\bf 5}, 3929 (1990).

\bibitem{mesons}
S.~Dalley and B.~van de Sande,
Phys.\ Rev.\ D {\bf 67}, 114507 (2003)
[arXiv:hep-ph/0212086].

\bibitem{glueballs} 
S. Dalley and B. van de Sande, 
Phys.\ Rev.\ Lett.\ {\bf 82}, 1088 (1999) 
[arXiv:hep-th/9810236]; 
Phys.\ Rev.\ D {\bf 62}, 014507 (2000) 
[arXiv:hep-lat/9911035].

\bibitem{sdmesons} S.~Dalley,
Phys.\ Rev.\ D {\bf 64}, 036006 (2001) 
[arXiv:hep-ph/0101318].

\bibitem{other}
M. V. Polyakov and C. Weiss, 
Phys. Rev. D {\bf 60}, 114017 (1999)
[arXiv:hep-ph/9902451];
\\
I. V. Anikin {\em et al}, 
Nucl. Phys. A {\bf 678}, 175 (2000)
[arXiv:hep-ph/9905332];
\\
A. Mukherjee, I. V. Musatov, H-C. Pauli, and A. V. Radyushkin,
Phys. Rev. D {\bf 67},  073014 (2003)
[arXiv:hep-ph/0205315];
\\
W. Broniowski and E. Ruiz Arriola, 
Phys. Lett. B {\bf 574}, 57 (2003)
[arXiv:hep-ph/0307198]; 
in {\em Lightcone Physics: Hadrons and Beyond}, p182 (Durham IPPP, 2003) 
[arXiv:hep-ph/0310048];
\\
L. Theussl, S. Noguera and V. Vento, 
Eur. Phys. J. A {\bf 20}, 483 (2004)
[arXiv:nucl-th/0211036];
\\ 
F. Bissey {\em et al.},  
Phys. Lett. B {\bf 587},  189 (2004)
[arXiv:hep-ph/0310184];
\\
M. Praszalowicz and A. Rostworowski, 
Acta Phys.Polon. B {\bf 34}, 2699 (2003)
[arXiv:hep-ph/0302269].

\bibitem{diehl} 
M. Diehl, 
Phys. Rept. {\bf 388}, 41 (2003)
[arXiv:hep-ph/0307382].

\end{thebibliography}
\end{document}